\documentstyle[11pt,aaspp4]{article}
\def\pn{\par\noindent}

\def\ltsima{$\; \buildrel < \over \sim \;$}
\def\ltsim{\lower.5ex\hbox{\ltsima}}
\def\gtsima{$\; \buildrel > \over \sim \;$}
\def\gtsim{\lower.5ex\hbox{\gtsima}}

\begin{document}

\title{Gamma-ray Bursts Produced by Mirror Stars\\
      Presented at XXVII ITEP Winter School, Snegiri, Feb. 16 -- 24, 1999\\
      Proceedings of the School to be published by Gordon \& Breach}

\author{Sergei Blinnikov
\affil{Institute for Theoretical and Experimental Physics,
 117259, Moscow, Russia; \\ email: sergei.blinnikov@itep.ru}
\affil{Sternberg Astronomical Institute, 119899, Moscow,
          Russia;
        \\ email: blinn@sai.msu.su}}

\begin{abstract}

I argue that cosmic Gamma-ray Bursts (GRB) may be produced by collapses
or mergers of stars made of `mirror' matter. The mirror neutrinos
(which are sterile for our matter) produced at these events
can oscillate into ordinary neutrinos. The annihilations or decays of
the latter create an electron-positron plasma and subsequent relativistic
fireball with a very low baryon load needed for GRBs. The concept of mirror
matter is able to explain several key problems of modern astrophysics:
neutrino anomalies, the missing mass, MACHO microlensing events and GRBs.
Thus this concept becomes very appealing and should be considered quite
seriously and attentively.

\end{abstract}


\section{Introduction}

The spectacular discovery of GRB afterglows allowed to measure the
redshift, and hence the distance to some of them. The energy output
up to $3.4 \times 10^{54}$ ergs $\approx 1.9 M_\odot c^2$, for GRB990123
(Kulkarni et al. 1999) poses extremely hard questions to theorists who try
to explain these superpowerful events. Even if a beaming is invoked, 
it can reduce the energy budget by two orders of magnitude, perhaps,
but this is still too high  for conventional models.

The extraordinary situation requires a revolutionary approach to the
modeling of GRBs. In this Lecture I suggest a scenario which seems to be
a bizarre one from the first glance, but in fact it has a reasonable
theoretical basis, and observational evidence in favour of this scenario is
ever growing: I believe, that observing
GRBs at cosmological distances we are witnessing catastrophic deaths of
stars made of the so called ``mirror'' matter.

\section {Problems in GRB modeling}

For general recent reviews on GRBs see e.g. Piran (1998a,b),
Tavani (1998) and Postnov (1999).

It is well known, that assuming high values of Lorentz factor $\Gamma$
of the GRB ejecta is necessary to solve the {\it compactness problem} 
(Guilbert, Fabian \& Rees 1983,  Paczy\'nski 1986, Goodman 1986,
Krolik \& Pier 1991,  Rees \&  M\'esz\'aros 1992, Piran 1996).  
The typical time-scale of the variability
of the gamma-ray emission $\Delta t\sim10^{-2}$~seconds implies the size
of the emitting region $R<c\Delta t$, as small as $\sim 10^3$~km. The   
enormous number of gamma photons in such a small volume should produce  
electron-positron pairs which make the emitting region optically thick. 
This conflicts with the observed nonthermal spectra unless one supposes
that the emitting region moves
towards  the observer at a relativistic speed with Lorentz factor $\Gamma$,   
then its size would be $\Gamma^2c\Delta t$, and the optical depth
correspondingly smaller. The low optical depth and
the ultrarelativistic motion requires that the fireball should be very
clean (not heavily contaminated with baryons), yet the models suggested
up to now are producing `dirty' fireballs.

E.g., the possibility of a GRB to appear during
a bare core collapse in a binary system was suggested by Dar et
al. (1992).  The latter model assumed a GRB to be a result of the
neutrino-antineutrino pair creation and annihilation
(Goodman et al. 1987) during the
accretion-induced collapse of a white dwarf in a close binary system.
Although the idea of neutrino annihilation is very compelling for
producing GRBs, the model should be rejected on the grounds of being too
contaminated by baryonic load, see e.g. Woosley (1993).

Another plausible way of forming GRBs at cosmological distances
involves binary neutron star merging (originally proposed by
Blinnikov et al. 1984; see more recent references and statistical
arguments in favour of this model in Lipunov et al. 1995).
However, as detailed hydrodynamical calculations currently
demonstrate, this mechanism also fails in producing powerful clean fireballs
(Janka and Ruffert 1996, Ruffert et al. 1997).
On the GRB models 
with a moderately high baryon load see Woosley (1993),
Ruffert \& Janka (1998), Klu\'zniak \& Ruderman (1998),
Fuller \& Shi (1998), Fryer \& Woosley (1998),
Popham, Woosley, \& Fryer (1998).
Vietri \& Stella (1998) and Spruit (1999) suggest models that probably have
a small contamination,  but it is unlikely to derive from them the huge
energy required by the most recent GRB observations.

A very interesting idea was put forward by Klu\'zniak (1998). He suggested
that the ordinary neutrinos can oscillate into sterile ones, go out to
the regions relatively free of baryons, and then convert back into ordinary
neutrinos. For this model the difficulty is the same: if the oscillation
length is too short than the baryon contamination is unavoidable. If it is
too long then a very small number of neutrinos will annihilate.

Here I point out to the possibility of dramatically extending
the latter model. The sterile neutrino are naturally produced by the
mirror matter during collapses or mergers of mirror stars, made of
mirror baryons. If they oscillate to ordinary neutrinos they do this
in the space practically free of ordinary baryons and can give birth
to a powerful gamma-ray burst.

\section {The concept of mirror matter}

The concept of the mirror particles stems from the idea of
Lee \& Yang (1956) who suggested the existence of new particles
with the reversed sign of the mirror asymmetry observed in
our world. Lee and Yang  believed that the new particles
(whose masses are degenerate with the masses of ordinary particles)
could participate in the ordinary interactions. Later, 
Kobzarev, Okun \& Pomeranchuk (1966) have shown that this conjecture was
not correct, and that the ordinary strong,
weak and electromagnetic interactions are forbidden for the mirror particles 
by experimental evidence, only gravity and super-weak interaction
is allowed for their coupling to the ordinary matter. But if they
really mirror the properties of ordinary particles, this means that there
must exist mirror photons, gluons etc., coupling the mirror fermions
to each other, like in our world. Thus the possibility of existence
of the mirror world was postulated first by
Kobzarev, Okun \& Pomeranchuk (1966), and the term ``mirror'' was coined
in that paper.
The particle mass pattern and particle interactions in the mirror world
are quite analogous to that in our world, but the two worlds interact
with each other essentially through gravity only.

Later the idea was developed in a number of papers, e.g.
Okun (1980), Blinnikov \& Khlopov (1983),
and the interest to it is revived recently in attempts to explain all
puzzles of neutrino observations Foot \& Volkas (1995),
Berezhiani \& Mohapatra (1995), Berezhiani et al. (1996), Berezhiani (1996).
It is shown in the cited
papers that a world of mirror particles can coexist with our, visible, world,
and some effects that should be observed are discussed.

It was shown by Blinnikov \& Khlopov (1983)
that ordinary and mirror matter are most likely well mixed on the
scale of galaxies, but not in stars, because of
different thermal or gasdynamic processes like SN shock waves which induce
star formation. It was predicted that star counts by HST
must reveal the deficit of local luminous matter if the mirror stars
do really exist in numbers comparable to ordinary stars and contribute
to the gravitational potential of galactic disk. 
Recent HST results Gould et al. (1997) show the reality of the luminous
matter deficit: e.g., instead of 500 stars expected from the Salpeter mass
function in the HST fields investigated for the range of absolute
visual magnitudes $ 14.5 < M_V < 18.5 $
only 25 are actually detected. It is found that the Salpeter slope
does not continue down to the hydrogen-burning limit but has a maximum
near $M \sim 0.6 M_\odot$, so lower mass stars do not contribute
much to the total luminous mass as was thought previously.
The total column density of the galactic disk, $\Sigma \approx 40
M_\odot {\rm pc}^{-2}$ is a factor of two lower than
published estimates of the dynamical mass of the disk Gould et al. (1997).
It should be
remembered that here we discuss a contribution of invisible stars to
the gravity of the galactic disk which has more to do with the local Oort
limit (see e.g. Oort 1958) than with the halo dark matter.
Other references on the subject see also in
Mohapatra \& Teplitz (1999).

Okun (1980), Blinnikov \& Khlopov (1983), Berezhiani (1996) have pointed out
that mirror objects can be observed by the effect of gravitational lensing.
After the discovery MACHO microlensing events, I have discussed their
interpretation as mirror stars at Atami meeting in 1996 (Blinnikov 1998).
Recently, this interpretation is developed by Foot (1999) and
Mohapatra \& Teplitz (1999).

The mirror world that interacts with ordinary matter exclusively via gravity
follows quite naturally from some models in superstring theory (closed
strings), but those models are too poor to be useful in our problem.
Especially interesting for explaining GRBs are the models that predict
the existence of a light sterile neutrino that can oscillate into
ordinary neutrino. The development of the idea can be traced from the
following references.

Foot et al. (1991), showed that the mirror symmetry is compatible with
the standard model of particle physics.
Here it was assumed that the neutrinos are massless, and
it was  shown that there are only two possible ways in addition
to gravity, that the mirror particles can interact with the 
ordinary ones, i.e. through photon-mirror photon mixing
[this had already been discussed independently and earlier
(in in a slightly different context) by Glashow (1985)]
and through Higgs-mirror Higgs mixing.

In the next paper Foot et al. (1992) have shown that if neutrinos have
mass then the mirror
idea can be tested by experiments searching for 
neutrino oscillations and can explain the solar neutrino problem.
The same idea can also explain the atmospheric
neutrino anomaly (recently confirmed by SuperKamiokande data), which
suggests that the muon neutrino is maximally mixed with another species.
Parity symmetry suggests that each of the three known neutrinos
is maximally mixed with its mirror partner (if
neutrinos have mass). This was pointed out by Foot (1994).
Finally, the idea is also compatible with the LSND experiment
which suggests that the muon and electron neutrinos oscillate
with small angles with each other, see Foot \& Volkas (1995).
Berezhiani \& Mohapatra (1995)
extended the latter work  to a bit different model with
parity symmetry spontaneously broken. In this model the mirror
particles have masses on all scales differing by a common factor from the
masses of their ordinary counterparts.

\section{The GRB model}

Now I am ready to formulate very briefly the scenario of my model.

If the properties of mirror matter are very similar to the
properties of particles of the visible world, then the events
like neutron star mergers, failed supernovae (with a collapse to a rotating
black hole) etc. must occur in the mirror world. These events can easily
produce sterile (for us) neutrino bursts with energies up to $10^{54}$ ergs,
and the duration and beaming of mirror neutrinos are organized naturally
like in the standard references given above. The neutrino
oscillations then take place which transform them at least partly
to ordinary neutrinos, but without the presence of big amounts
of visible baryons. Some number of ordinary baryons is needed, like
$10^{-5} M_\odot$ (Piran 1998b) for producing standard afterglows etc.
This number is easily accreted by mirror stars during their life from
the uniform ordinary interstellar matter (cf. Blinnikov and Khlopov 1983).
The oscillation length required in this scenario must be less than the size
of the system (10 -- 100 km) multiplied by the number of scatterings of the
mirror neutrinos in the body of mirror neutron star, $\sim 10^5$.

A variety of properties of GRBs can be explained as suggested by
Klu\'zniak \& Ruderman (1998) for ordinary matter.

Taking into account magnetic moment of standard neutrinos can help in
producing a larger variety of GRB variability due to neutrino interaction
with the turbulent magnetic field inevitably generated in the fireball.
This is good for temporal features similar to the observed fractal or
scale-invariant properties found in gamma-ray light curves of GRB (Shakura et
al. 1994;  Stern and Svensson 1996). Another extension of the model is
possible if heavier neutrinos can decay into lighter ones 
producing photons directly (see e.g. Jaffe  \& Turner 1997).

\section{Conclusion: arguments in favour of mirror matter models}

Summarizing, here are the arguments in favour of the propose scenario.

\begin{enumerate}

\item  The mirror matter is aesthetically appealing, because it restores
the parity symmetry of the world (at least partly).
\item It allows to explain neutrino anomalies.
\item It explains the missing mass in the Galaxy disk,
 and in some models the  Dark matter in general.
\item It explains MACHO microlensing events
\item For GRBs it provides the model with the low baryon load
\item The available baryon load on the scale of the mass of a small
      planet is exactly what is needed for fireball models.
\item All host galaxies for OT of GRBs are
strange ones. This may be an indication for the gravitational interaction
of the ordinary galaxy with the mirror one in which it can be immersed.

\end{enumerate}
\pn
{\bf Acknowledgements.} I am very grateful to Lev Okun, Konstantin Postnov,
Ilya Tipunin, Mikhail Prokhorov, Darja Cosenco, Aleksandra Kozyreva,  
Elena Sorokina for stimulating discussions and assistance,
and to Robert Foot for interesting correspondence. 

%\begin{references}

%\end{references}

\small

\begin{thebibliography}{}

\bibitem{Ber95} Berezhiani Z.G., Mohapatra R.N., 1995, Phys.Rev. D52, 6607
\bibitem{Beretal96} Berezhiani Z.G., Dolgov A.D., Mohapatra R.N., 1996,
 Phys. Lett. B375, 26;
\bibitem{Ber96} Berezhiani Z.G., 1996, Acta Phys.Polon. B27,  1503
\bibitem{blinatam}
 Blinnikov S.I., 1998,
  presented at "Baryonic Matter in the Universe and Its
  Spectroscopic Studies" (November 22 - 24, 1996, Atami, Japan),
 astro-ph/9801015
 %Title: A quest for weak objects and for invisible stars
 %Authors: S. I. Blinnikov (ITEP, Moscow)
\bibitem{BlKh} Blinnikov S.I., Khlopov, M.Yu., 1983, Astron.Zh. 60,
 632 [translation: Sov.Astron. 27, 371]
\bibitem{bnpp}
 Blinnikov S.I., Novikov I.D., Perevodchikova T.V.,
 Polnarev A.G., 1984, PAZh, 10,  422 [translation: SvA Letters, 10, 177]
\bibitem{darknp}
 Dar A.,  Kozlovsky B.Z., Nussinov S.,  Ramaty, R.,  1992, ApJ, 388, 164
\bibitem{Foot}  Foot R., 1994, Mod. Phys. Lett. A9, 169
\bibitem{Foot99}Foot R., 1999, astro-ph/9902065
 %Title: Have mirror stars been observed?
\bibitem{FLH} Foot R., Lew H., Volkas R.R., 1991, Phys. Lett. B272, 67
\bibitem{FLV} Foot R., Lew H., Volkas R.R., 1992, Mod. Phys. Lett. A7,
 2567
\bibitem{FootV} Foot R., Volkas R.R., 1995,  Phys.Rev. D52, 6595
\bibitem[Fryer \& Woosley 1998]{1998ApJ...502L...9F} Fryer, C. L. \& 
 Woosley, S. E. 1998, \apjl, 502, L9
\bibitem[Fuller \& Shi 1998]{fullershi} Fuller G. M., Shi X.,  1998,
 %Supermassive Objects as Gamma-Ray Bursters,
 ApJ, 502, L5
\bibitem{Glashow} Glashow S.L., 1985, Phys. Lett. B167, 35
\bibitem{GBF} Gould A., Bahcall J.N., Flynn C. ApJ 482 (1997) 913
% (astro-ph/9611157)
\bibitem{goodman} Goodman J., 1986, ApJ, 308, L47
\bibitem[Goodman, Dar, \& Nussinov 1987]{goodmdr} Goodman J.,
 Dar A.,  Nussinov S., 1987, 
 %Neutrino annihilation in Type II supernovae,
 ApJ, 314, L7
\bibitem{Guilbert83} Guilbert P.W., Fabian A.C., Rees M.J., 1983,
 MNRAS, 205, 593
\bibitem{JT} Jaffe A., Turner M. S.  1997, Phys.Rev. D55, 7951
%Title: Gamma Rays and the Decay of Neutrinos from SN1987A
\bibitem[Janka \& Ruffert 1996]{1996A&A...307L..33J} Janka, H. -T. \&
 Ruffert, M. 1996, \aap, 307, L33
\bibitem{LY} Lee T.D., Yang C.N., Phys.Rev. 104 (1956) 256
\bibitem[Klu\'zniak 1998]{1998ApJ...508L..29K} Klu\'zniak, W.
 1998, \apjl, 508, L29
\bibitem[Klu\'zniak \& Ruderman 1998]{1998ApJ...505L.113K}
 Klu\'zniak, W., Ruderman, M. 1998, \apjl, 505, L113
\bibitem{KOP} Kobzarev I., Okun L., Pomeranchuk I., 1966,
 Yadernaya Fizika 3, 1154 [translation: Sov.J.Nucl.Phys. 3, 837]
\bibitem{krolik91} Krolik J.H., Pier E.A., 1991, ApJ, 373, 277
\bibitem{LPPPJ} Lipunov V.M., Postnov K.A., Prokhorov M.E., Panchenko I.E.,
 Jorgensen H.E., 1995, ApJ, 454, 593
\bibitem{MohTep} Mohapatra R.N., Teplitz V.L., 1999, astro-ph/9902085
 %Title: Mirror Matter MACHOs
\bibitem{Okun} Okun L.B., 1980, ZhETF 79, 694 [translation:
 Sov. Phys. JETP 52, 351 ]
\bibitem{Oort} Oort J.H., in Stellar Populations, ed. O'Connel,
 Vatican, 1958, p. 145
\bibitem{pacz} Paczy\'nski B., 1986, ApJ, 308, L43
\bibitem{Piran96} Piran T.,  1996, in {\it Unsolved Problems in Astrophysics}
 Eds. Bahcall J. N., Ostriker J. P., Princeton University Press, p.343
\bibitem{PiranUP} Piran T., 1998a, to appear in the proceedings
 of the Fifth Conference on Underground Physics, TAUP97, astro-ph/9801001
\bibitem{PiranReview} Piran T., 1998b, to appear in Physics Reports,
 astro-ph/9810256 %     Title: Gamma-Ray Bursts and the Fireball Model
\bibitem{powoo98} Popham R., Woosley S.E., Fryer C.,
   submitted to The Astrophysical Journal (astro-ph/9807028)
\bibitem{Postnov} Postnov K.A., 1999, GRBs: A Review,
 submitted to Uspekhi Fiz. Nauk
\bibitem{ReesMes} Rees M.J.,  M\'esz\'aros P., 1992, MNRAS, 258, 41P
\bibitem[Ruffert et al. 1997]{rufjts} Ruffert M., Janka H.-T., Takahashi K.,
 Schaefer G. 1997,  A\&A, 319, 122
 %Coalescing neutron stars - a step towards physical models.
 %II. Neutrino emission, neutron tori, and gamma-ray bursts.,
\bibitem[Ruffert \& Janka 1998]{1998A&A...338..535R} Ruffert, M. \& Janka, 
 H. -T.  1998, \aap, 338, 535
 %Colliding neutron stars. Gravitational waves,
 %neutrino emission, and gamma-ray bursts
Title: Gamma Rays and the Decay of Neutrinos from SN1987A
Authors: A. Jaffe (CITA), M. S. Turner (Chicago, Fermilab)
Comments: 17 pages, 6 postscript figures, uses revtex, epsf.sty. Submitted to PRD. Also available at file://ftp.cita.utoronto.ca/cita/jaffe/papers/pvoflux.ps.gz
Journal-ref: Phys.Rev. D55 (1997) 7951-7959
\bibitem[Spruit 1999]{1999A&A...341L...1S} Spruit, H. C. 1999, \aap, 341, L1 
\bibitem{SS} Stern B.E.,  Svensson R. 1996, ApJ, 469, L109
\bibitem{SPS} Shakura N.N., Prokhorov M.E., Shakura, N.I.,  1994, Astronomy
 Lett., 22, 137
\bibitem[Tavani 1998]{Tavani} Tavani M., 1998
    Astrophysical Letters and Communications, in press,
  astro-ph/9812422
% Title: Gamma-Ray Bursts: the Four Crises
% Adapted from a paper presented
% at the 3rd INTEGRAL Workshop: The Extreme Universe, Taormina 14-18 Sept. 1998,
\bibitem[Vietri 1998]{Vietri} Vietri M., Stella L.,  1998, \apjl, submitted,
 astro-ph/9808355
% A gamma ray burst with small contamination
\bibitem{woo93}Woosley S.E., 1993, ApJ, 405, 273
\end{thebibliography}
\end{document}